# Variable Coupling Strength of Silicene on Ag(111)

FENG Bao-Jie(冯宝杰)[1], LI Wen-Bin(黎文彬)[1], QIU Jing-Lan(邱静岚)[1], CHENG Peng(程鹏)[1], CHEN Lan(陈岚)[1], WU Ke-Hui(吴克辉)[1,2**]

*1 Institute of Physics, Chinese Academy of Sciences, Beijing 100190, China*
*2 Collaborative Innovation Center of Quantum Matter, Beijing 100871, China*
Corresponding author. Email: khwu@iphy.ac.cn

**We performed a scanning tunneling microscopy and spectroscopy (STM/STS) study on the electronic structures of √3×√3-silicene on Ag(111). We find that the coupling strength of √3×√3-silicene with Ag(111) substrate is variable at different regions, giving rise to notable effects in experiments. These evidences of decoupling or variable interaction of silicene with the substrate are helpful to in-depth understanding of the structure and electronic properties of silicene.**

1. Introduction

Theoretical explorations for layered group IV materials can be dated back to two decades ago [1], and recently, the monolayer silicon with $sp^2$ hybridization, namely silicene, has been predicted to be stable [2-4]. Similar to graphene, the band structure of silicene exhibits linear energy-momentum dispersion with high Fermi velocity ($10^5$~$10^6$ m/s), forming six Dirac cones at the K points in the reciprocal space. Interestingly, due to the larger spin-orbit coupling than that in graphene, silicene is expected to open a much larger energy gap at the Dirac point, leading to detectable quantum spin Hall effect [4]. Furthermore, by applying electric and magnetic field, the band structure of silicene is controllable among different phases, such as quantum spin Hall phase, quantum anomalous Hall phase, valley polarized metal phase, and so on [5]. Far more than these, silicene is expected to have a lot of other novel properties [6-10].

On the other hand, experimental realization of silicene has been successful only until recently [11-14]. Silicene was reported to epitaxially grow on Ag(111), ZrB$_2$, and Ir(111), but predominately on Ag(111). On Ag(111), silicene can form various structure, such as 3×3, (√7×√7)R19.1º, (√3×√3)R30º (simplified as √3-silicene) with respect to silicene-1×1 [11,13,15], while most structures are still under debate. For the 3×3 silicene, angle resolved photoemission spectroscopy (ARPES) measurements observed a linear energy-momentum dispersion [11]. However, the combined STM experiments and first-principles calculations have shown that the Dirac cone of the 3×3 silicene is seriously modified by strong hybridization of silicene with the substrate [16], and further ARPES measurements have confirmed the absence of Dirac cones in the 3×3 silicene [17]. For the √3-silicene, STM experiments have observed pronounced quasiparticle interference (QPI) patterns [18], which is absent in any other silicene structures on Ag(111). By analyzing the QPI patterns, a linear energy-momentum dispersion has also been derived [18], and quasiparticles have defined chirality arising from the pseudospin structure of silicene [19]. In addition, below 40 K, the √3-silicene will undergo a structural phase transition [20] and show a superconducting-like gap at the Fermi level [21]. For the multilayer √3-silicene, ARPES experiments have observed a linear dispersion, which is an evidence of Dirac fermions in √3-silicene [22].

In spite of the fruitful experimental results of √3-silicene, its structure is still under debate. Some experiments and theoretical calculations proved the existence of monolayer √3-silicene [13,20] while others claimed that the thinnest √3-silicene is bilayer [23-25]. Recently, some researchers even claim that the √3-silicene is (√3×√3)-Ag on top of ultrathin Si(111) film [26,27]. However, the metallic nature of the surface states in Si(111)√3×√3-Ag [28,29] contradict with the quasiparticle chirality and hexagonal warping observed previously [19]. Anyway, the richness of novel properties of √3-silicene makes it a promising material that needs further exploration.

Here, by performing STM/STS experiments on √3-silicene, we find that the coupling strength of the film to the Ag(111) substrate can be different in different regions. In the strongly coupled region, clear moiré patterns have been observed in the topographic images, in contrast to the weakly coupled region where the moiré pattern is hardly observable. Furthermore, we observe a remarkable suppression of the characteristic flat band of silicene at 0.9 V in the strongly coupled regions, which can also be successfully explained by the different coupling strength of silicene with Ag(111).

## 2. Experimental details

Experiments were carried out in a home-built low temperature STM with a base pressure of $6\times10^{-9}$ Pa. Single crystal Ag(111) sample was cleaned by repeated cycles of Ar$^+$ ion sputtering and annealing at 600°C. Silicene was obtained by evaporating silicon from a silicon wafer (≈1000°C) onto Ag(111). The temperature of Ag(111) is kept at approximately 550 K during the growth process and the deposition rate of silicon was 0.05 ML/min (here 1 monolayer refers to the atomic density of the ideal silicene sheet). Our STM and STS experiments were carried out at 78 K and 5 K, and the bias voltage was applied to the tip, *i.e.* positive bias voltage corresponds to the occupied states.

## 3. Results and Discussion

At sufficiently high substrate temperature and appropriate silicon coverage, a well-ordered (√3×√3)R30° honeycomb superstructure of silicene forms on Ag(111), as shown in Fig.1(a). Unlike the coexistence of various structure of silicene at low growth temperature [13,23,24], here we obtain only one structure, the √3-silicene, on Ag(111). This √3-silicene will undergo a structural phase transition at 5 K (Fig.1(b)), as reported previously [20]. The structure of √3-silicene can be explained by a spontaneous ultra-bulking of silicene driven by weak van der Waals interaction between silicene and Ag(111) [20]. The structural model of √3-silicene is shown in Fig.1(d). The two triangles are mirror symmetric √3 superstructure which will flip-flop quickly at high temperature. But at sufficiently low temperature (below 40 K), the two mirror symmetric structures can be distinguished, forming the domain boundaries shown in Fig.1(c).

Interestingly, we noticed frequently that some regions appear slightly lower than the neighboring region, even though they are within exactly the same terrace and the same domain. As shown in Fig. 2(a) and (b), there are two different regions labeled A and B which are in the same continuous √3-silicene domain. However, the height is different for the two regions. Region A is 64 pm lower than region B at 1.0 V and 40 pm lower than region B at -0.6 V. This height difference is much lower than the height of one atomic layer. So it could not be attributed to the underlying step edges. Another prominent difference is that region B exhibits clear moiré patterns at certain voltages, as shown in Fig.2(b), while the moiré pattern in region A in the topographic images is obscure. From the FFT map of region B, we can clearly see the spots of √3 lattice and the moiré pattern. A simple calculation show that the orientation of silicene with respect to Ag(111) is 28º which is close to the values 30º measured by directly comparing the orientation of the silicene domain and the neighboring bare Ag(111) substrate. Interestingly, the FFT map of region A also show faint spots of moiré pattern which is the same as that in region B. This fact clearly shows that the orientation of silicene with respect to Ag(111) substrate is the same in the two regions. The appearance of moiré patterns in √3-silicene also gives strong evidence that the √3-silicene is not the Si(111)-√3×√3-Ag.

When cooled down to 5 K, the √3-silicene undergoes a structural phase transition, as shown in Fig.3. As can be found from the image, the domain boundaries are continuous across the two regions, which shows that the phase transition is the intrinsic properties of the √3-silicene, irrespective of the differences of the two regions.

We further performed STS measurements on the two regions. The dI/dV curve of √3-silicene is characterized by a sharp peak at 0.9 V below the Fermi level which originates from a flat band in silicene [30]. From Fig.3(c), we find that the dI/dV curves in region A and B are almost the same. But the intensity of the peak at 0.9 V is slightly lower in region B than that in region A, which means the flat band is suppressed in region B. At other energies, the intensity of the dI/dV curves is almost the same. The blue curve in Fig.3(c) is obtained by subtracting curve B from A, which shows a peak at 0.9 V, while at other energies from -1.5 V to 1.5 V, it is almost zero. This can be easily seen from the dI/dV maps in Fig.4. At 0.9 V, the intensity of region A is weaker than that of region B, while at other voltages, such as 0.7 V and -1.4 V, the two regions have almost the same intensity, as shown in Fig.4(c). It needs to note that the QPI patterns are also visible in region B and the wavelength is the same in the two regions, so linear dispersion can be derived from the QPI patterns. This means that the Dirac fermion characteristic is preserved in region B [20].

The appearance of two different regions of √3-silicene can be explained by different coupling strength of silicene with the substrate. At region B, the coupling is stronger, leading to a smaller distance between the silicene layer and the Ag(111) substrate. The stronger coupling with the substrate produces clearer moiré patterns, [31] which can perfectly explain our results. It is worth to note that in an STM experiment on graphite, Li *et al*. [32] found some decoupled regions with larger distance between the surface layer and the underlying graphite. This could serve as additional evidence that may prove the variable coupling strength of silicene with Ag(111) is possible.

The different coupling strength of √3-silicene with Ag(111) can also explain the suppression of the flat band in silicene. In previous theoretical calculations, it has been pointed out that there is a flat band in the gap at the Dirac point in monolayer √3-silicene [20]. Considering the charge transfer from Ag(111) to silicene, this flat band may correspond to the one observed at 0.9 eV below the Fermi level. According to this scenario, the flat band is mainly contributed by the $p_z$ atomic orbitals of the three low buckled Si atoms around the high buckled atom [20]. So a weakening of the flat band will occur when the interaction of silicene with the Ag(111) substrate is stronger. This is because the interaction of silicene with Ag(111) will involve the contribution of $p_z$ atomic orbitals of silicon atoms and thus will weaken its contribution to the flat band. For the bilayer √3-silicene model, theoretical calculations also show a flat band at 1 eV below the Fermi level [25], which is another possible origin of the flat band at 0.9 eV. Based on this bilayer model, in region B, the stronger interaction of silicene with Ag(111) will also suppress its contribution to the flat band and result in the weakening of the peak at 0.9 eV.

4. Conclusion

In summary, we performed STM/STS experiments to study the electronic properties of √3-silicene. We found that the coupling strength is variable in different regions. The difference in the coupling strength results in experimentally observable effects, such as the height difference, moiré pattern strength, and the intensity of the flat band, despite of the common features in the electronic properties. Such variable coupling strength may be useful for tuning the properties of silicene in further applications.

**Acknowledgement:** This work was supported by the MOST of China (Grants No. 2012CB921703, 2013CB921702), the NSF of China (Grants No. 11334011, 91121003), and the Strategic Priority Research Program of the Chinese Academy of Sciences (Grants No. XDB07000000).

**References:**

[1] Takeda K and Shiraishi, K 1994 Phys. Rev. B **50** 14916

[2] Guzman-Verri G G and Voon L C L Y 2007 Phys. Rev. B **76** 075131

[3] Cahangirov S, Topsakal M, Aktürk E, Sahin, H and Ciraci, S. 2009 Phys. Rev. Lett. **102**

236804

[4] Liu C-C, Feng W and Yao Y 2011 Phys. Rev. Lett. **107** 076802

[5] Ezawa M 2012 Phys. Rev. Lett. **109** 055502

[6] An X-T, Wang, Y-Y and Liu J-J and Li S-S 2012 New J. Phys. **14** 083039

[7] Stille L, Tabert, C J and Nicol E J 2012 Phys. Rev. B **86** 195405

[8] Tsai W-F, Huang C-Y, Chang T-R, Lin H, Jeng H-T and Bansil A 2013 Nat. Commu. **4** 1500

[9] Kikutake K, Ezawa M and Nagaosa N 2013 Phys. Rev. B **88** 115432

[10] Linder J and Yokoyama T 2014 Phys. Rev. B **89** 020504(R)

[11] Vogt P, De Padova P, Quaresima C, Avila J, Frantzeskakis E, Asensio M C, Resta A, Ealet B and Le Lay G 2012 Phys. Rev. Lett. **108** 155501

[12] Fleurence A, Friedlein R, Ozaki T, Kawai H, Wang Y and Yamada-Takamura Y 2012 Phys. Rev. Lett. **108** 245501

[13] Feng B, Ding Z, Meng S, Yao Y, He X, Cheng P, Chen L and Wu K 2012 Nano Lett. **12** 3507

[14] Meng L, Wang Y, Zhang L, Du S, Wu R, Li L, Zhang Y, Li G, Zhou H, Hofer W A and Gao H-J 2013 Nano Lett. **13** 685

[15] Enriquez H, Vizzini S, Kara A, Lalmi B and Oughaddou H 2012 J. Phys.: Condens. Matter. **24** 314211

[16] Lin C-L, Arafune R, Kawahara K, Kanno M, Tsukahara N, Minamitani E, Kim Y, Kawai M and Takagi N 2012 Phys. Rev. Lett. **110** 076801

[17] Tsoutsou D, Xenogiannopoulou E, Golias E, Tsipas P and Dimoulas A 2013 Appl. Phys. Lett. **103** 231604

[18] Chen L, Liu C-C, Feng B, He X, Cheng P, Ding Z, Meng S, Yao Y and Wu K 2012 Phys. Rev. Lett. **109** 056804

[19] Feng B, Li H, Liu C-C, Shao T-N, Cheng P, Yao Y, Meng S, Chen L and Wu K 2013 ACS Nano **7** 9049

[20] Chen L, Li H, Feng B, Ding Z, Qiu J, Cheng P, Wu K and Meng S 2013 Phys. Rev. Lett. **110** 085504

[21] Chen L, Feng B and Wu K 2013 Appl. Phys. Lett. **102** 081602

[22] De Padova P, Vogt P, Resta A, Avila J, Razado-Colambo I, Quaresima C, Ottaviani C, Olivieri

B, Bruhn T, Hirahara T, Shirai T, Hasegawa S, Asensio M C and Le Lay G 2013 Appl. Phys. Lett. **102** 163106

[23] Arafune R, Lin C-L, Kawahara K, Tsukahara N, Minamitani E, Kim Y, Takagi N and Kawai M 2013 Surf. Sci. **608** 297

[24] Vogt P, Capiod P, Berthe M, Resta A, De Padova P, Bruhn T, Le Lay G and Grandidier B 2014 Appl. Phys. Lett. **104** 021602

[25] Guo Z-X and Oshiyama A 2014 Phys. Rev. B **89** 155418

[26] Shirai T, Shirasawa T, Hirahara T, Fukui N, Takahashi T and Hasegawa S 2014 Phys. Rev. B **89** 241303(R)

[27] Mannix A J, Kiraly B, Fisher B L, Hersam M C and Guisinger N P 2014 ACS Nano **7** 7538

[28] Crain J N, Altmann K N, Bromberger C and Himpsel F J 2012 Phys, Rev. B **66** 205302

[29] Takahashi K, Azuma J and Kamada M 2012 J. Electron. Spectrosc. Relat. Phenom. **185** 547

[30] Feng Y, Feng B, Xie Z, Li W, Liu X, Liu D, Zhao L, Chen L, Zhou X and Wu K 2014 Chin. Phys. Lett. **31** 127303

[31] Kobayashi K 1994 Phys. Rev. B **50** 4749

[32] Li G, Luican A and Andrei E Y 2009 Phys. Rev. Lett. **102** 176804

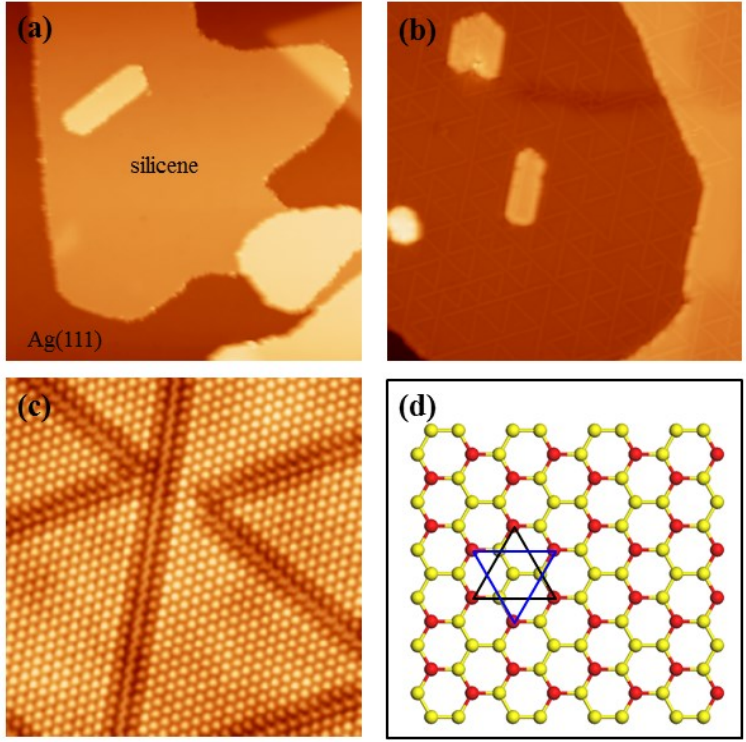

**Figure 1**: (color) (a)-(c) STM topographic images. Size: (a) 100×100 nm$^2$, (b) 100 nm×100 nm, (c) 20×20 nm$^2$. Bias voltage: (a) 1.2 V, (b) -2.0 V, (c) -0.8 V. (a) is obtained at 78 K and (b) and (c) are obtained at 5 K. (d) Structure model of the √3-silicene. The black and blue triangles illustrate the mirror symmetric √3 superstructures of silicene which will flip-flop at high temperature.

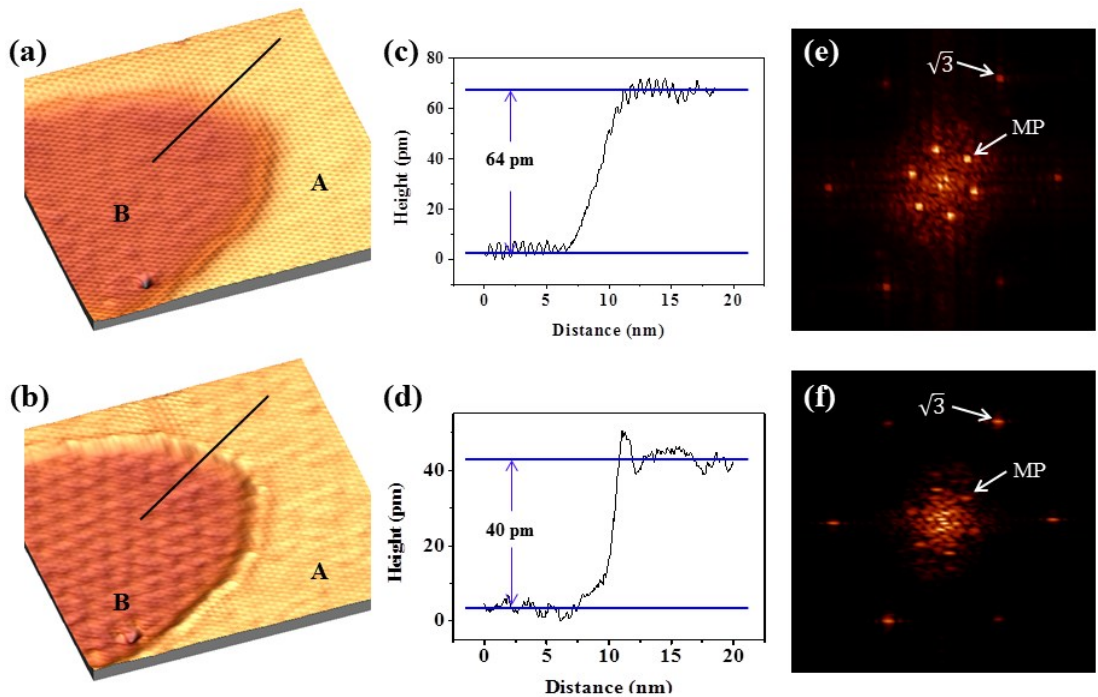

**Figure 2:** (color) (a) STM topographic image (30×30 nm$^2$, 1.0 V) of a continuous silicene sheet

with two different regions, labeled A and B. (b) STM topographic image at the same area as (a) but with a different bias voltage -0.6 V. (c) and (d) Line profiles along the black line in (a) and (b), respectively. (e) and (f) Fast Fourier transformation images of region B and A in (b), respectively. The STM images in (a) and (b) are both obtained at 78 K.

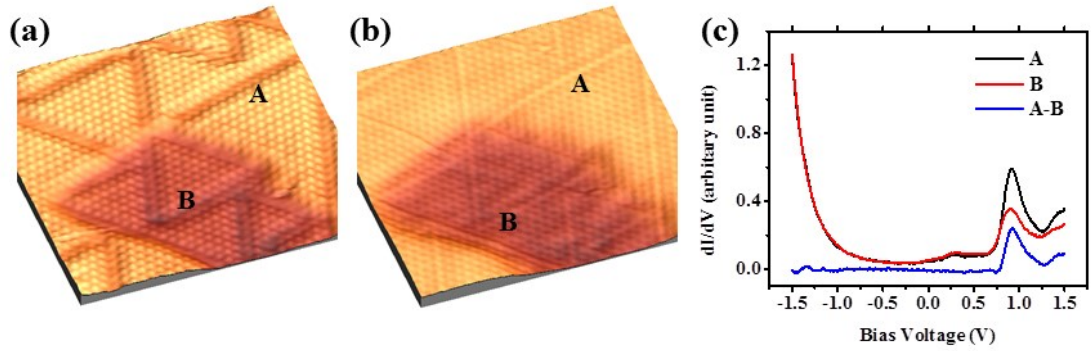

**Figure 3:** (color) (a) STM topographic images (30×30 nm$^2$, -0.9 V) with similar two regions as shown in Fig.2. (b) STM topographic images of the same area as (a) with a different bias voltage 1.0 V. (c) Typical dI/dV spectra at region A (black) and B (red), respectively. The blue line is obtained by subtraction of B from A. All the date are obtained at 5 K.

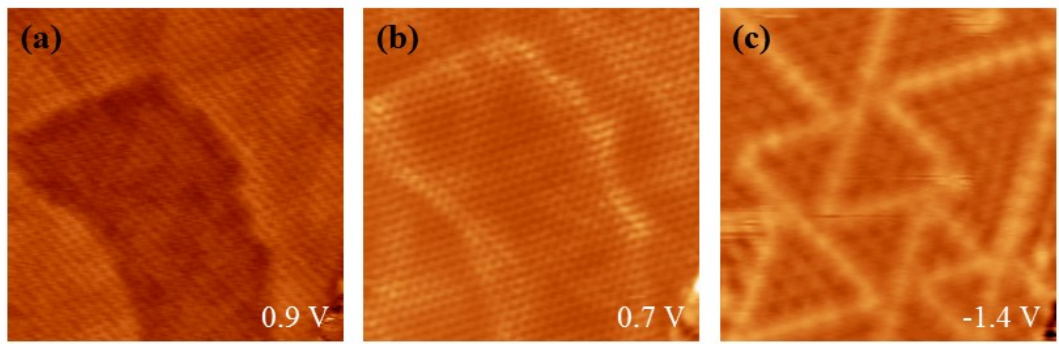

**Figure 4:** (color) (a)-(c) dI/dV maps at the same area as the topographic images in Fig.3. The bias voltages are 0.9 V, 0.7 V and -1.4 V respectively.